\documentclass[aip, apl, superscriptaddress, reprint, showpacs]{revtex4-2}

\usepackage{amsfonts}
\usepackage{amsmath}
\usepackage{chemformula}
\usepackage{amssymb, bm, mathrsfs}
\usepackage{graphicx, epstopdf, color}
\usepackage{dcolumn}
\usepackage{bm}
\usepackage{soul}
\usepackage{appendix}
\usepackage[colorlinks=true,citecolor=blue]{hyperref}

\begin{document}

\title{Injection-Limited and Space-Charge-Limited Conduction in Wide Bandgap Semiconductors with Velocity Saturation Effect}

\author{Kok Wai Lee}
\affiliation{Science and Math, Singapore University of Technology and Design, Singapore 487372}

\author{Yee Sin Ang}
\email{yeesin\_ang@sutd.edu.sg}
\affiliation{Science and Math, Singapore University of Technology and Design, Singapore 487372}

\begin{abstract}

Carrier conduction in wide bandgap semiconductors (WBS) often exhibits velocity saturation at the high-electric field regime. How such effect influences the transition between contact-limited and space-charge-limited current in a two-terminal device remains largely unexplored thus far. Here, we develop a generalized carrier transport model that includes contact-limited field-induced carrier injection, space charge, carrier scattering and velocity saturation effect. The model reveals various transitional behaviors in the current-voltage characteristics, encompassing Fowler-Nordheim emission, trap-free Mott-Gurney (MG) SCLC and \emph{velocity-saturated SCLC}. Using GaN, 6H-SiC and 4H-SiC WBS as examples, we show that the velocity-saturated SCLC completely dominates the high-voltage ($10^2 \sim 10^4$ V) transport for typical sub-$\mu$m GaN and SiC diodes, thus unravelling velocity-saturated SCLC as a central transport mechanism in WBG electronics.  

\end{abstract}

\maketitle
Space-charge-limited current (SCLC) is the maximum current density which can be injected across a diode of channel length, \(D\) subjected to a bias voltage, \(V\) \cite{10.1063/1.4978231}.  Very generally, SCLC can be described by the scaling law:
\begin{equation}
    J \propto \frac{V^\alpha}{L^\beta}
\end{equation}
whereby the scaling exponents $\alpha$ and $\beta$ are determined by the material properties and device geometry \cite{Zhang-Ang-Garner}. 
In a one-dimensional (1D) diode, SCLC can be broadly classified into ballistic and collisional types. The former occurs in vacuum whereby SCLC is governed by the Child-Langmuir (CL) law \cite{Child, Langmuir} with $(\alpha, \beta) = (3/2, 2)$. The latter occurs in solids such as semiconductors and insulators in which carrier scattering effects are present. Depending on the defect characteristics and device geometry \cite{Zubair-Fatima}, the scaling factors $(\alpha,\beta)$ can take the form of: (i) $(3,2)$ for the cases of trap-free, trap-filled, and traps with discrete energy level (i.e. Mott-Gurney (MG) \cite{Mott}, Lampert \cite{Lampert} and Rose \cite{Rose} models); (ii) $(l+1, 2l+1)$ for a shallow exponential distribution of traps whereby $l$ is a parameter related to trap characteristics and temperature (i.e. Mark–Helfrich (MH) model \cite{Mark-Helfrich}); (iii) $(3/3)$ and $(l+1,l+1)$ for trap-free (trap-filled) and trap-limited ultrathin diode, respectively \cite{Grinberg, kee2022analytical}; and (iv) $(1/2,1/2)$ for two-dimensional (2D) Dirac semiconductors \cite{Ang-Zubair}.
Due to the sensitive dependence of SCLC current-voltage-length scaling on the material properties and device geometry, SCLC has been widely used as an effective \emph{diagnostic tool} to probe the material properties (e.g. carrier mobility, trap density) \cite{Rohr,Corre}, and has been extensively used to investigate organic \cite{Buhl} and inorganic semiconductors \cite{Karunagaran, Partain}, metal halide perovskites \cite{Duijnstee, Corre, Almora-Osbel, Almora-Miravet}, dielectrics and wide bandgap (WBG) semiconductors \cite{Zhang-Lin, Shi, Zhang-Xiang, Gaska, Adivarahan}.

Modelling the transitions between various transport processes into SCLC is essential for device design and characterization of both vacuum as well as solid-state diodes and transistors. 
Various transitional behaviours have been previously uncovered, including the transitions from contact-limited to space-charge-limited in the ballistic \cite{Lau} and collisional \cite{Darr-Loveless, Chua} regimes, the inclusion of thermionic effect \cite{Darr-Darr}, and the peculiar absence of SCLC in 2D materials \cite{Chua_absence}. 
It should be noted that prior transitional models \cite{Darr-Loveless, Chua, Chua_absence} typically assume a \emph{constant} carrier mobility, which is only approximately valid at the low-electric field regime.
For most solids, the carrier mobility becomes field-dependent at higher field.
In particular, in wide bandgap semiconductors (WBS), the field-dependent mobility can be described by the Caughey-Thomas mobility model \cite{Caughey} where the the charge carrier velocity saturates towards a maximum drift velocity, known as the \emph{saturation velocity} ($v_s$) \cite{Ridley, Shishir, Roccaforte, Chen, Smithe, Soboleva, Woods-Robinson}, which arises from carrier scattering effects such as optical phonon scattering \cite{Schenk}. 
How velocity saturation affects and/or competes with SCLC (note that both occurs at high fields) and a general transport model encompassing contact-limited field emission, SCLC and velocity saturation effect remain open questions thus far.

\begin{figure*}[ht]
    \centering
    \includegraphics[width = \textwidth]{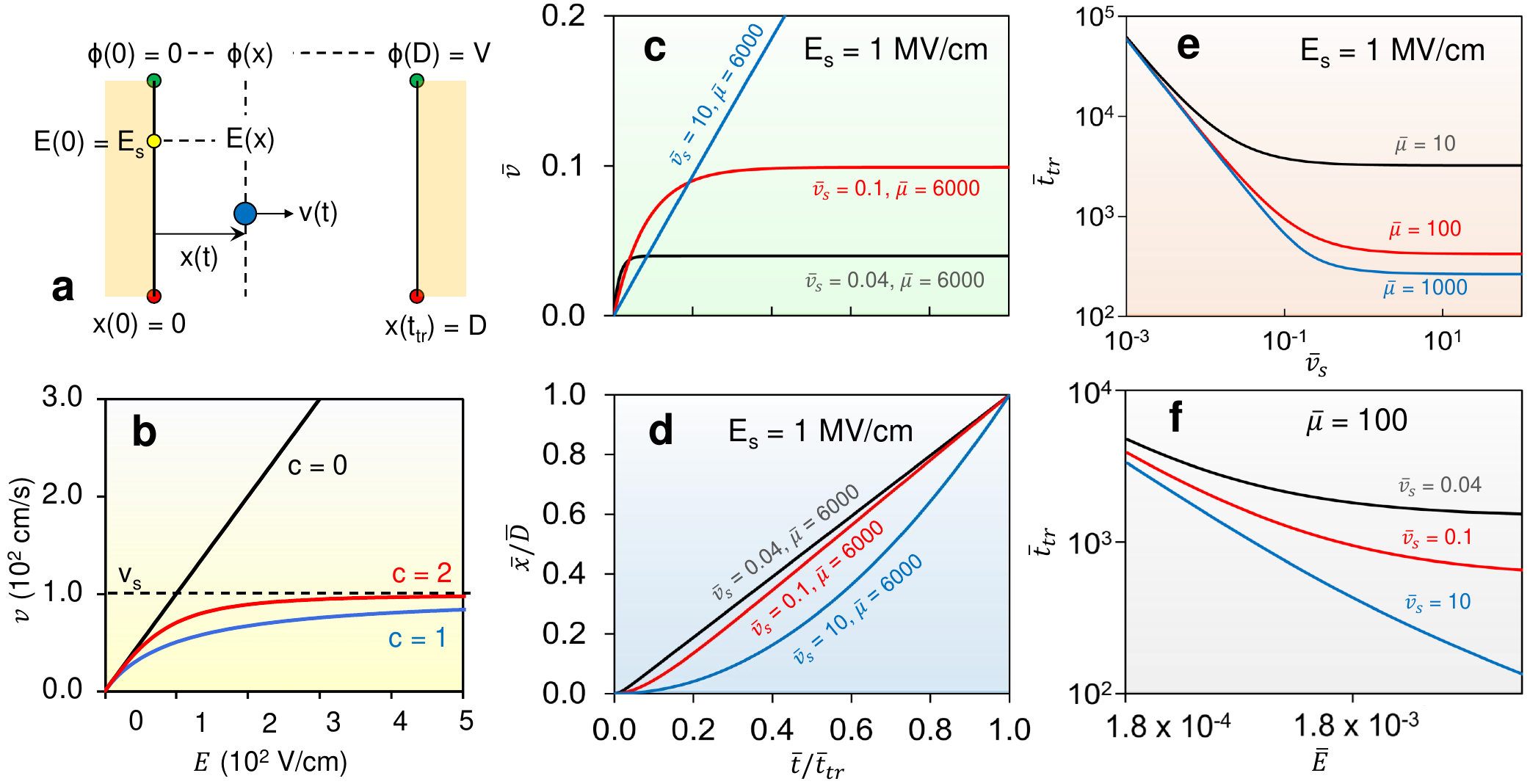}
    \caption{(a) Schematic diagram of the \(1\)D charge carrier transport. (b) \(v-E\) plot at \(v_s = 100\) cm/s for \(c = 0,1\) and \(2\). (c) \(\Bar{v} -\Bar{t}/\Bar{t}_{tr}\) and (d) \(\Bar{x}/\Bar{D}-\Bar{t}/\Bar{t}_{tr}\) plots at \(E_s = 1\) MV/cm for different \(\Bar{v}_s\). (e) \(\Bar{t}_{tr} - \Bar{v}_{s}\) plot at  \(E_s = 1\) MV/cm for different \(\Bar{\mu}\). (f) \(\Bar{t}_{tr} - \Bar{E}\) plot at \(\Bar{\mu} = 100\) for different \(\Bar{v}_s\). The device length is $D = 100$ nm. }
    \label{fig: Figure 1} 
\end{figure*}   

In this work, we develop a generalized transport model that covers both contact-limited and space-charge-limited conduction regimes in the presence of carrier scattering and velocity saturation effects. By combining the carrier equation of motion, Caughey-Thomas mobility model and Poisson's equation, the current-voltage (\(J-V\)) characteristics for both electron and hole transports are obtained, which enable the rich transitional behaviors between field emission (FE), trap-free Mott-Gurney SCLC (SCLC$_\text{MG}$) and the velocity-saturated SCLC (SCLC$_\text{VS}$) to be elucidated. We construct the nexus plots to illustrate various transitions under varying material/device parameters. Using the archetypal WBS of gallium nitride (GaN), and silicon carbide (SiC) polytypes of 6H-SiC and 4H-SiC, we show that the current-voltage curve transits directly from the contact-limited FE to SCLC$_\text{VS}$ while the SCLC$_\text{MG}$ commonly observed in dielectrics and insulators is completely absent, thus suggesting the importance of velocity saturation effect in WBS-based devices operating at high voltages.

The model developed here shall provide a theoretical basis for the modelling and analysis of WBS-based devices -- a key building block of high-voltage/high-temperature electronics much sought-after for electric vehicles \cite{matallana2019power} and harsh environment applications \cite{bader2020prospects}, and can also be used to investigate the effect of velocity saturation on vacuum diode with residual gas \cite{Darr-Darr,Darr-Loveless} as well as the breakdown phenomenon in micro/nanogap devices \cite{loveless2017universal, wang2022transitions}. 

We consider the one-dimensional (along the \(x\)-direction) charge carrier transport across a planar diode with the cathode located at \(x = 0\) and the anode located at \(x = D\) biased at a voltage, \(V\) with respect to the cathode as shown in Fig. \ref{fig: Figure 1}(a). The position and velocity of the injected carrier are $x(t)$ and $v[x(t)]$ respectively. The local electric field at $x$ is $E[x(t)]$. The carrier injected from the cathode into the diode bulk is assumed to have a zero initial (\(t = 0\)) velocity, \(v\) and such injection is driven by the surface electric field, \(E_{s}\), giving rise to the set of initial conditions \(x(0) = 0\), \(v(0) = 0\) and \(E(0) = E_{s}\).
\par
We first model the electrostatics and carrier transport by coupling Poisson's equation with the current density equation, \(J = n[x(t)]ev[x(t)]\) with \(n[x(t)]\) and \(e\) being the position-dependent carrier density and electric charge constant respectively. Thus, we obtain $d^2\phi(x)/dx^2 = J/\epsilon v$ whereby \(\phi(x)\) is the electric potential and \(\epsilon\) is the dielectric constant of the diode channel \cite{Fowler}. By using the electric field $E[x(t)] = d\phi[x(t)] / dx$ and carrier velocity $v[x(t)] = dx(t)/dt$ relations, integration with respect to $x$ yields
\begin{equation}
    E[x(t)] = \frac{d\phi[x(t)]}{dx} = E_{s} + \frac{J(E_s)t}{\epsilon}
\end{equation}
where we have used the initial condition of $E[x(t)] = E_s$ at $t=0$. The current flowing in the diode is determined by that injected from the cathode which can be described by the Fowler-Nordheim (FN) law, 
\begin{equation}
    J(E_s) = A_\text{FN} E_s^2 \exp\left( - \frac{B_\text{FN}}{E_s} \right)
\end{equation}
whereby $E_s$ is the surface electric field at the cathode, while $A_\text{FN}$ and $B_\text{FN}$ are the first and second FN constants. 

Next, we derive the equation of motion (EOM) of a charge carrier traversing across the diode in the presence of carrier scattering or collisional effect. Such effect is captured by a friction term in the EOM \cite{Darr-Loveless}, i.e.
\begin{equation}
    m \frac{dv[x(t)]}{dt} = e E[x(t)] - \alpha(E[x(t)]) v[x(t)]
\end{equation}
whereby $\alpha(E[x(t)])$ is a `friction-coefficient-like' term related to the strength of the collision, and is generally field-dependent. At steady state, Eq. (4) yields $v[x(t)] = [e/\alpha(E[x(t)])] E[x(t)]$ where the carrier mobility can be written as $\mu \equiv e/\alpha[(E[x(t)])]$ to yield the familiar expression of $v[x(t)] = \mu E[x(t)]$. 
The carrier mobility is approximately field-independent at the low-field regime but acquires significant field-dependence at the high-field regime. 
Since SCLC is a high-field phenomenon, the inclusion of field-dependent carrier mobility is particularly important in the modelling of SCLC in solid-state diodes. \par

\begin{figure*}[ht]
    \centering
    \includegraphics[width = \textwidth]{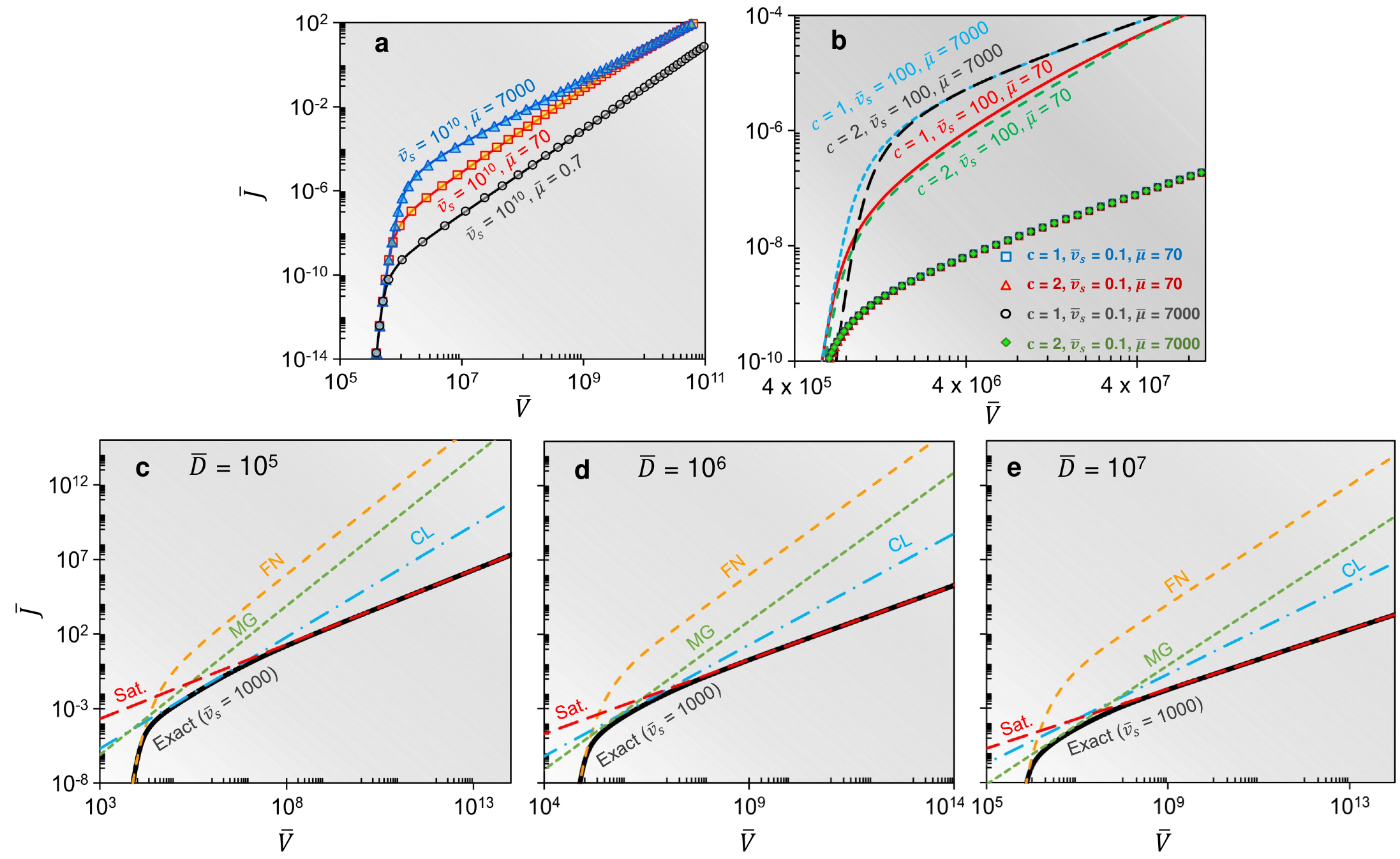}
    \caption{(a) Comparison between \(\Bar{J} - \Bar{V}\) characteristic curves obtained by our model with \(c = 1\) (markers) and 
    obtained by Darr et al \cite{Darr-Loveless} (solid lines) for \(\Bar{\mu} = 0.7, 70\) and \(7000\) at \(\Bar{D} = 10^7\). (b) Comparison between  \(\Bar{J} - \Bar{V}\) characteristic curves for \(c = 1\) and \(2\) at \(\Bar{D} = 10^7\) for different \(\Bar{\mu}\) and \(\Bar{v}_s\). \(\Bar{J} - \Bar{V}\) characteristic curves  at \(\Bar{\mu} = 700\) for (c) \(\Bar{D} = 10^5\), (d) \(\Bar{D} = 10^6\) and (e) \(\Bar{D} = 10^7\).}
    \label{fig: Figure 2}
\end{figure*}    

To incorporate the field-dependent collisional effect into the EOM, we consider the Caughey-Thomas carrier mobility model \cite{Caughey}:

\begin{equation}
    \mu(E) = \frac{\mu_0 E}{\left[1 + \left(\frac{\mu_0 E}{v_s}\right)^c\right]^\frac{1}{c}}
\end{equation}
\\
whereby $\mu_0$ is the low-field carrier mobility and we suppress the argument of $E[x(t)]$ for notational simplicity. At the high-electric field regime of $E \gg v_s/\mu_0$, the carrier velocity approaches the saturation velocity \(v_{s}\) as shown in Fig. \ref{fig: Figure 1}(b). The coefficient \(c\) is an experimental fitting parameter generally equals to 1 (2) for holes (electrons). Nevertheless, the actual value may differ from 1 or 2 and is typically determined empirically from experimental data. For instance, \(c\) = 2 is a suitable fitting value to be used to determine the saturation velocity of graphene on \ch{SiO2} \cite{Dorgan} whereas \(c\) ranging from 2.8 to 4.8 are suitable fitting values to be used to determine a range of values for the saturation velocity of synthetic monolayer \ch{MoS2} \cite{Smithe}. \(v_s\) is material-dependent which is influenced by various factors such as temperature and doping density \cite{Moon, Chandrasekar}. The value of \(v_s\) varies across different materials and could range from \(10^5\) cm/s to \(10^7\) cm/s. For instance, amorphous indium gallium zinc oxide (a-IGZO) has a saturation velocity of \(3.8 \times 10^5\) cm/s \cite{Su} whereas gallium arsenide (GaAs) has a much higher saturation velocity of \(1.2 \times 10^7\) cm/s. 
\begin{widetext}

Combining Eqs. (2), (4) and (5), we obtain the field-dependent EOM, 
\begin{equation}
   \frac{dv[x(t)]}{dt} = \frac{e}{m}\left(E_s + \frac{Jt}{\varepsilon}\right) - \frac{e}{m\mu_0}v[x(t)]\left[1 + \frac{\mu_0}{v_s}\left(E_s + \frac{Jt}{\varepsilon}\right)^c\right]^{1/c}
\end{equation}
whereby \(m\) is the electron effective mass. For a given $E_s$, the injection current, $J(E_s)$ is determined from Eq. (3). For any value of $c$, Eq. (6) can  be solved numerically to yield the carrier velocity and position profiles, $v(t)$ and $x(t)$ respectively. The transit time, $t_\text{tr}$ -- time taken for a charge carrier to traverse across the diode -- is obtained using the boundary condition, $x(t_\text{tr}) = D$. Integrating Eq. (6) yields
\begin{equation}
    V(E_s) = \frac{1}{e} \frac{mv[x(t_\text{tr})]^2}{2} + \frac{1}{m\mu_0} \int_0^{t_\text{tr}} \left[ 1 + \frac{\mu_0}{v_s} \left(E_s + \frac{Jt}{\epsilon}\right)^c \right]^{1/c}v[x(t)]^2 dt 
\end{equation}
whereby $V(E_s)$ is the bias voltage and we have used $dx = v dt$ and Eq. (2) to obtain Eq. (7). Eq. (7) can be used to construct the current-voltage characteristics together with Eq. (3). For $c=1$, $v(t)$ and $x(t)$ can be solved for analytically from Eq. (6), yielding %
    \begin{eqnarray}
    \Bar{v}(\Bar{t}) &=& \Bar{v}_s \biggl\{1 - \exp\left[\frac{-\Bar{t}(2\Bar{v}_s + 2\Bar{\mu} \Bar{E} + \Bar{\mu} \Bar{J}\Bar{t})}{2\Bar{v}_s \Bar{\mu}}\right]\biggl\}  + \frac{\Bar{v}_s^\frac{3}{2}}{\Bar{\mu}}\sqrt{\frac{2}{\Bar{J}}} \biggl\{\exp\left[\frac{-\Bar{t}(2\Bar{v}_s + 2\Bar{\mu} \Bar{E} + \Bar{\mu} \Bar{J}\Bar{t})}{2\Bar{v}_s \Bar{\mu}}\right]F\left(\psi\right) - F\left(\gamma\right)\biggl\}
    \end{eqnarray}
    \begin{eqnarray}
    \Bar{x}(\Bar{t}) &=& \Bar{v}_s\Bar{t} + \Bar{v}_s^\frac{3}{2}\sqrt{\frac{\pi}{2\Bar{J}}}\exp\left(\psi^2\right)\left[\text{erf}\left(\psi\right) - \text{erf}\left(\gamma\right)\right] + \frac{\pi \Bar{v}_s^2}{2\Bar{\mu} \Bar{J}}\:\text{erfi}\left(\psi\right)\left[-\:\text{erf}\left(\psi\right) + \text{erf}\left(\gamma\right)\right] \nonumber \\
    && - \frac{\Bar{v}_s^\frac{3}{2}}{\Bar{\mu}}\sqrt{\frac{\pi}{2\Bar{J}}}\biggl\{-\frac{\Bar{v}_s(\Bar{v}_s + \Bar{\mu} \Bar{E})^2 {}_2F_2\left(1,1; \frac{3}{2},2; -\psi^2\right)}{\Bar{\mu}^2 \sqrt{2\pi}(\Bar{v}_s\Bar{J})^\frac{3}{2}} + \frac{\Bar{v}_s\left[\Bar{v}_s + \Bar{\mu}(\Bar{J}\Bar{t} + \Bar{E})\right]^2 {}_2F_2\left(1, 1; \frac{3}{2}, 2; -\gamma^2\right)}{\Bar{\mu}^2 \sqrt{2\pi}(\Bar{v}_s\Bar{J})^\frac{3}{2}}\biggl\}   
    \end{eqnarray}
    whereby 
    \begin{eqnarray}
        \psi = \frac{\Bar{v}_s + \Bar{\mu} \Bar{E}}{\Bar{\mu} \sqrt{2\Bar{J}\Bar{v}_s}} ; &
        \gamma = \frac{\Bar{v}_s + \Bar{\mu}(\Bar{E} + \Bar{J}\Bar{t})}{\Bar{\mu} \sqrt{2\Bar{J}\Bar{v}_s}}.
    \end{eqnarray}
    Here $F(x)$ is Dawson's integral \cite{Dawson}, \text{erf}(\(x\)) and \text{erfi}(\(x\)) are the error and imaginary error functions respectively and \({}_2F_2\)(\(a_1,a_2;\:b_1, b_2;\:z)\) is the generalized hypergeometric function of order \(p = q = 2\). 
    The normalisation terms \cite{Darr-Darr, Darr-Loveless} are defined as: \(\phi = \phi_{0}\Bar{\phi}\), \(J = J_{0}\Bar{J}\), \(x = x_{0}\Bar{x}\), \(t = t_{0}\Bar{t}\), \(\mu = \mu_{0}\Bar{\mu}\), \(E = E_{0}\Bar{E}\), \(v = v_{0}\Bar{v}\), \(v_{s} = v_{0}\Bar{v}_{s}\) whereby \(\phi_{0} = e\epsilon_{0}^2/mA_{FN}^2\), \(J_{0} = A_{FN}B_{FN}^2\), \(x_{0} = e\epsilon_{0}^2/mA_{FN}^2B_{FN}\), \(t_{0} = \epsilon_{0}/A_{FN}B_{FN}\), \(\mu_{0} = e\epsilon_{0}/mA_{FN}B_{FN}\), \(E_{0} = B_{FN}\) and \(v_{0} = x_{0}/t_{0}\).  
    In the large-$v_s$ regime, Eqs. (8) and (9) reduce to the collisional model of Darr et al \cite{Darr-Loveless}. 
    More interestingly, at large $J$, Eq. (7) becomes $V \to (v_s / 2m\epsilon) t_\text{tr}^2$ where $t_\text{tr} \to D/v_s$ approaches the time taken to traverse across the channel by a constant saturation velocity $v_s$. This leads to the velocity-saturated SCLC (SCLC$_\text{VS}$) of $J_{VS} = 2\epsilon v_s V / D^2$ \cite{sze2021physics} which, as demonstrated below, dominates the high-voltage transport in WBS-based devices.


 
\end{widetext}

\begin{figure*}[ht]
        \centering
        \includegraphics[width = 0.7\textwidth]{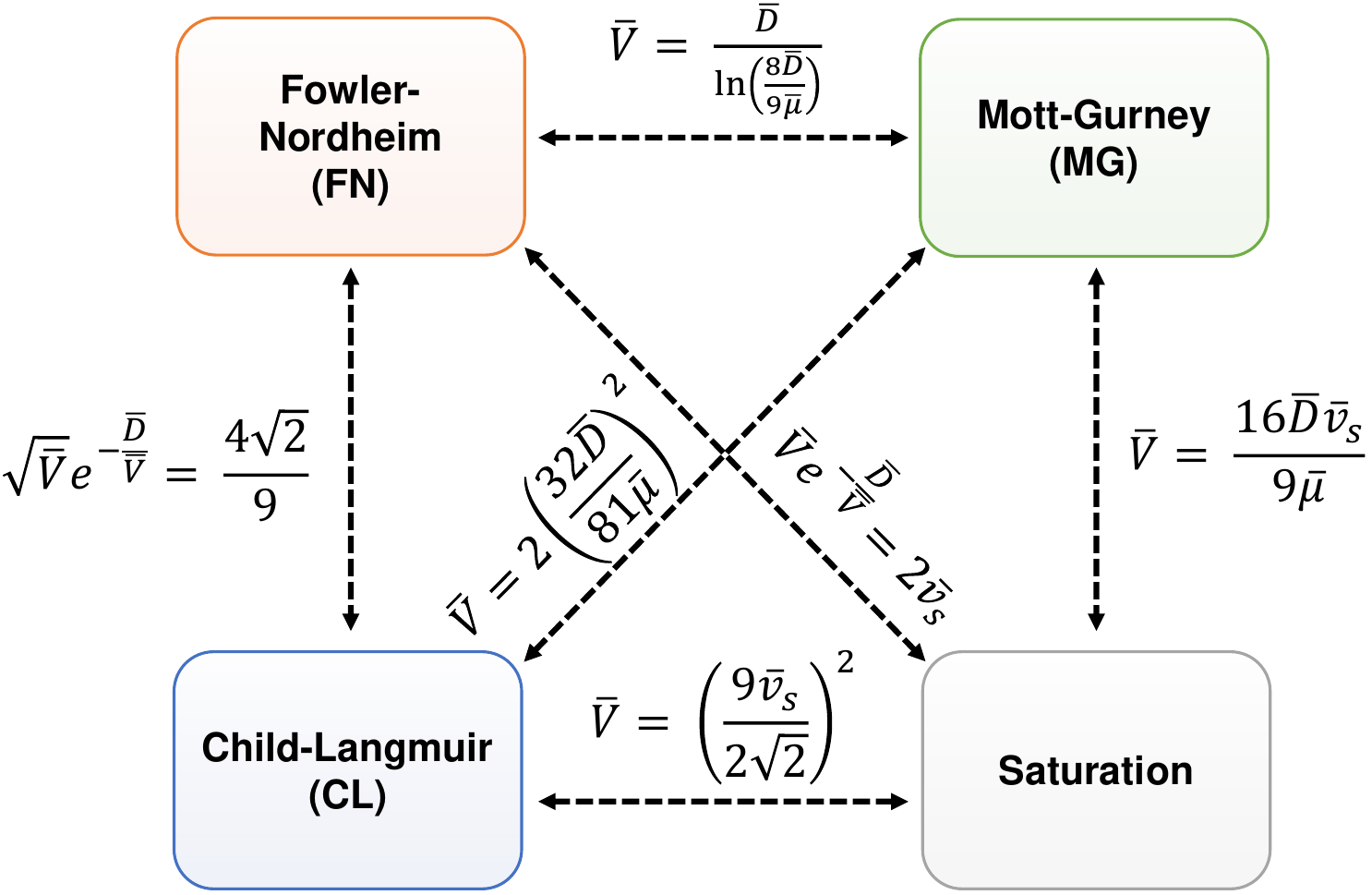}
        \caption{Schematic diagram of all possible transitions between the four transport regimes and their respective transition boundary equations.} 
        \label{fig: Figure 3}
\end{figure*}

\begin{figure*}[ht]
        \centering
        \includegraphics[width = \textwidth]{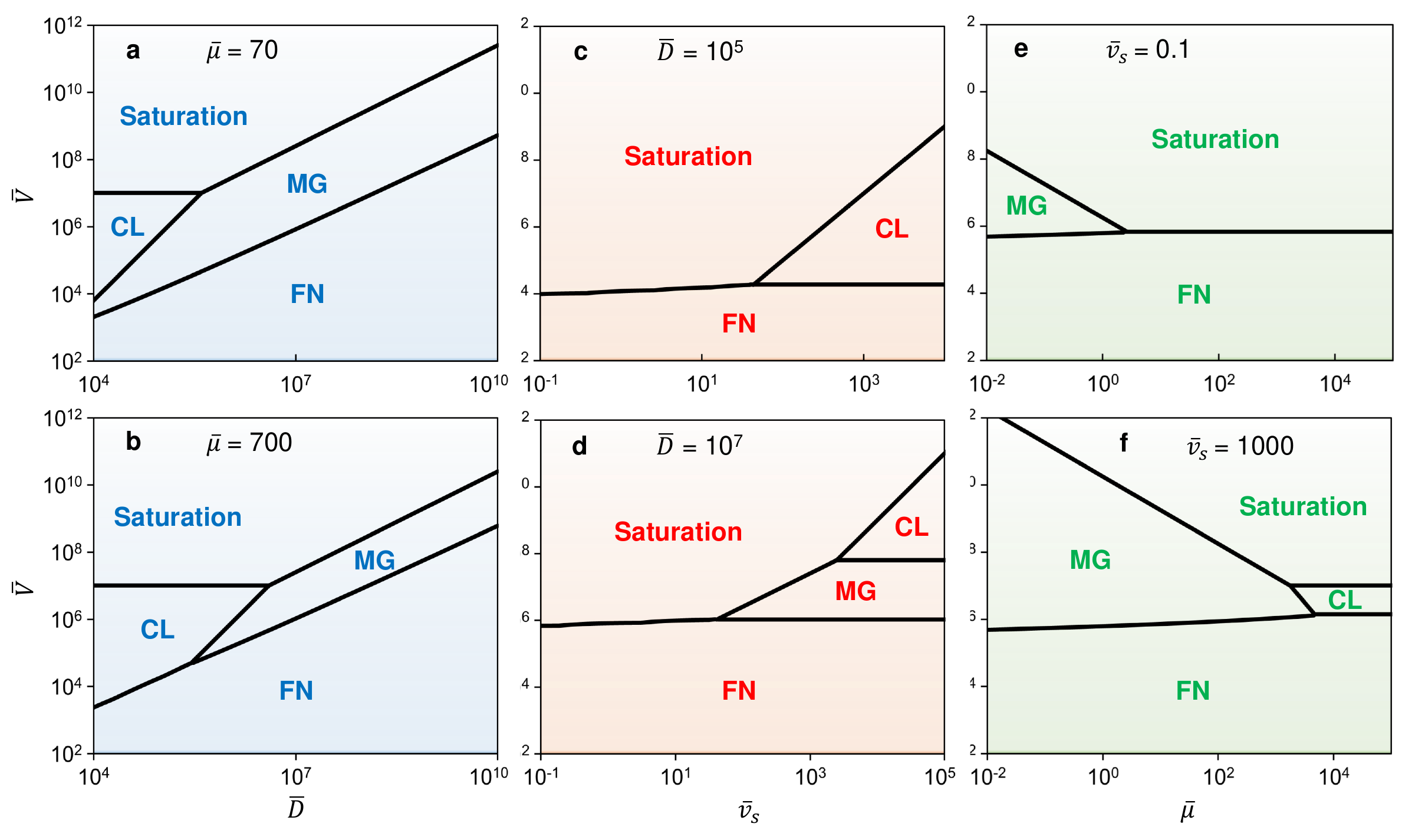}
        \caption{\(\Bar{V} - \Bar{D}\) nexus plot at \(\Bar{v}_s = 1000\) for (a) \(\Bar{\mu} = 70 \), (b) \(\Bar{\mu} = 700 \); \(\Bar{V} - \Bar{v}_{s}\) nexus plot at \(\Bar{\mu} = 700\) for (c) \(\Bar{D} = 10^5\), (d) \(\Bar{D} = 10^7\) and  \(\Bar{V} - \Bar{\mu}\) nexus plot at \(\Bar{D} = 10^7\) for (e) \(\Bar{v}_s = 0.1\), (f) \(\Bar{v}_s = 1000\).}
        \label{fig: Figure 4}
\end{figure*}

Fig. \ref{fig: Figure 1}(b) shows the carrier velocity as a function of \(E\) for three different values of $c$. For $c$ = 0, the velocity saturation effect is absent and a \(v-E\) linear relationship is obtained. For both $c = 1$ and $2$, the curves approach the saturation velocity  of $100$ cm/s as $E$ increases. In this case, the curve of $c = 2$ reaches $v_s$ faster when compared to that of $c = 1$. Although the velocity profiles in Fig. \ref{fig: Figure 1}(b) are noticeably different between $c=1$ and $2$, the current-voltage characteristics are not sensitively influenced by $c$ (see also, Fig. \ref{fig: Figure 2}(b) below). Thus, we focus on $c=1$ throughout Figs. \ref{fig: Figure 1} (c) to (f). 

Figs. \ref{fig: Figure 1}(c) and \ref{fig: Figure 1}(d) depict the normalised velocity and position profiles, respectively. For the case of $\Bar{v}_s = 10$, the velocity does not saturate as compared to those of $\Bar{v}_s = 0.1$ and $0.04$ within the time frame shown in Fig. \ref{fig: Figure 1}(c) since it takes a longer time to reach a larger $\Bar{v}_s$. 
Because the case of $\Bar{v}_s = 0.04$ rapidly saturates, the corresponding $\Bar{x}$ time-profile [Fig. \ref{fig: Figure 1}(d)] is nearly a positive-sloped straight line, which is akin to that of a \emph{constant-velocity} motion. In contrast, for the large-$\Bar{v}_s$ case, the $\Bar{x}$ exhibits parabolic-like temporal profile which is akin to the case of \emph{constant-acceleration} motion.


Fig. \ref{fig: Figure 1}(e) shows the transit time ($\Bar{t}_{tr}$), i.e. the time taken for a charge carrier to travel the entire channel from $x=0$ to $x=D$, as a function of $\Bar{v}_s$. Generally, $\Bar{t}_{tr}$ exhibits contrasting behaviours between low and high $\Bar{v}_s$: (i) $\Bar{t}_{tr}$ decreases rapidly with increasing $\Bar{v}_s$ at smaller $\Bar{v}_s$ and (ii) the decreasing trend saturates at larger $\Bar{v}_s$. Such contrasting behaviours can be explained as followed. At low $\Bar{v}_s$, the carrier velocity is severely limited by the saturation effect. Increasing $\Bar{v}_s$ raises the `speed limit' of the carrier. The $\Bar{t}_{tr}$ is thus reduced with increasing $\Bar{v}_s$. As $\Bar{v}_s$ continues to increase to a larger value, the carriers reach the end of the channel before their velocity can reach $\Bar{v}_s$. In this case, the transport is no longer $\Bar{v}_s$-limited and $\Bar{v}_s$ has minimal influence on $\Bar{t}_{tr}$, thus leading to the plateau at the high-$\Bar{v}_s$ regime.
We also note that an increasing $\Bar{\mu}$ appreciably decreases the $\Bar{t}_{tr}$ since larger mobility corresponds to an overall larger carrier velocity and hence a shorter transit time. 
The $\Bar{t}_{tr}$ decreases nearly monotonously with an increasing $\Bar{E}$ [Fig. \ref{fig: Figure 1}(f)] since larger $\Bar{E}$ leads to a larger driving force for accelerating the carrier, thus resulting in an overall larger carrier velocity and, similarly, a shorter transit time. Furthermore, when the carrier velocity is strongly $\Bar{v}_s$-limited (i.e. low $\Bar{v}_s$), the overall transit time is larger since the carrier velocity is rapidly capped by the lower $\Bar{v}_s$.

By setting $\Bar{v}_s$ to be $10^{10}$ so that the transport is nearly unaffected by the saturation effect, the $\Bar{J}-\Bar{V}$ characteristics obtained for $c = 1$ from our model is in excellent agreement with that obtained by a previous model of Darr et al \cite{Darr-Loveless} which does not include the velocity saturation effect [Fig. \ref{fig: Figure 2}(a)]. 
The velocity saturation effect on the $\Bar{J}-\Bar{V}$ characteristics only becomes more prominent at lower $v_s$, leading to a velocity-saturated SCLC transport regime (i.e. SCLC$_\text{VS}$) which is not captured by previous transitional model \cite{Darr-Loveless}. To investigate the significance of the fitting parameter, $c$, on the transport behaviour, the $\Bar{J}-\Bar{V}$ characteristics for $c = 1$ and $2$ are plotted in Fig. \ref{fig: Figure 2}(b). Apart from slight differences at low voltage, the $\Bar{J}-\Bar{V}$ characteristics almost overlap between $c = 1$ and $c=2$. As the $\Bar{J}-\Bar{V}$ characteristics are not sensitively influenced by $c$, electron and hole conductions are expected to yield similar $\Bar{J}-\Bar{V}$ characteristics as well as the transitional behaviours. We thus focus on $c=1$ in the following section.
\par

Figs. \ref{fig: Figure 2}(c) to \ref{fig: Figure 2}(e) show the $\Bar{J}-\Bar{V}$ characteristics with different $\Bar{D}$. 
Short-channel [Fig. \ref{fig: Figure 2}(c)] and long-channel [Fig. \ref{fig: Figure 2}(e)] devices exhibit contrasting $\Bar{J}-\Bar{V}$ characteristics. 
At the short channel regime, a distinct CL-type SCLC (SCLC$_\text{CL}$) occurs at the intermediate voltage regime. In this case, the carrier ballistically traverses across the entire channel before collisions can occur, thus leading to the vacuum-like SCLC$_\text{CL}$. In contrast, at the long-channel limit, the dominance of carrier scattering effect leads to a distinctive SCLC$_\text{MG}$ at the intermediate regime. 
Interestingly, with intermediate-channel length ($\bar{D} = 10^6$), the moderate-voltage regime exhibits both SCLC$_\text{MG}$ and SCLC$_\text{CL}$ scaling behaviours, thus simultaneous presence of both vacuum-like and solid-like characteristics of such device. 
Ultimately, the transport universally converges towards SCLC$_\text{VS}$ at the high-voltage limit for all $\Bar{D}$. 

We now discuss the transitions between various transport processes by constructing the \emph{nexus plots}, which can be obtained by equating their respective current density equations (see Fig. \ref{fig: Figure 3} for schematic illustration of various transitional regimes). Fig. \ref{fig: Figure 4} reveals the rich transitional behaviours of the $\Bar{J}$-$\Bar{V}$ characteristics across various device/material parameter spaces. 
For all channel lengths [Figs. \ref{fig: Figure 4}(a) and (b)], the SCLC$_\text{VS}$ is the ultimate transport process at high-voltage regime. At low-to-intermediate voltage, SCLC$_\text{CL}$ (SCLC$_{MG}$) dominates the short-channel (long-channel) devices -- a behaviour consistent with the $\Bar{J}$-$\Bar{V}$ characteristics in Fig.  \ref{fig: Figure 2}. For device with intermediate length, a more complex transition of FN-SCLC$_\text{MG}$-SCLC$_\text{CL}$-SCLC$_\text{VS}$ with increasing voltage can occur, suggesting a solid-like to vacuum-like transport transition as observed in Fig. \ref{fig: Figure 2}(d). Such effect is especially obvious in a low-mobility system [Fig. \ref{fig: Figure 4}(a)] due to the prominence of carrier scattering and collisional effects.

\begin{figure*}[ht]
        \centering
        \includegraphics[width = \textwidth]{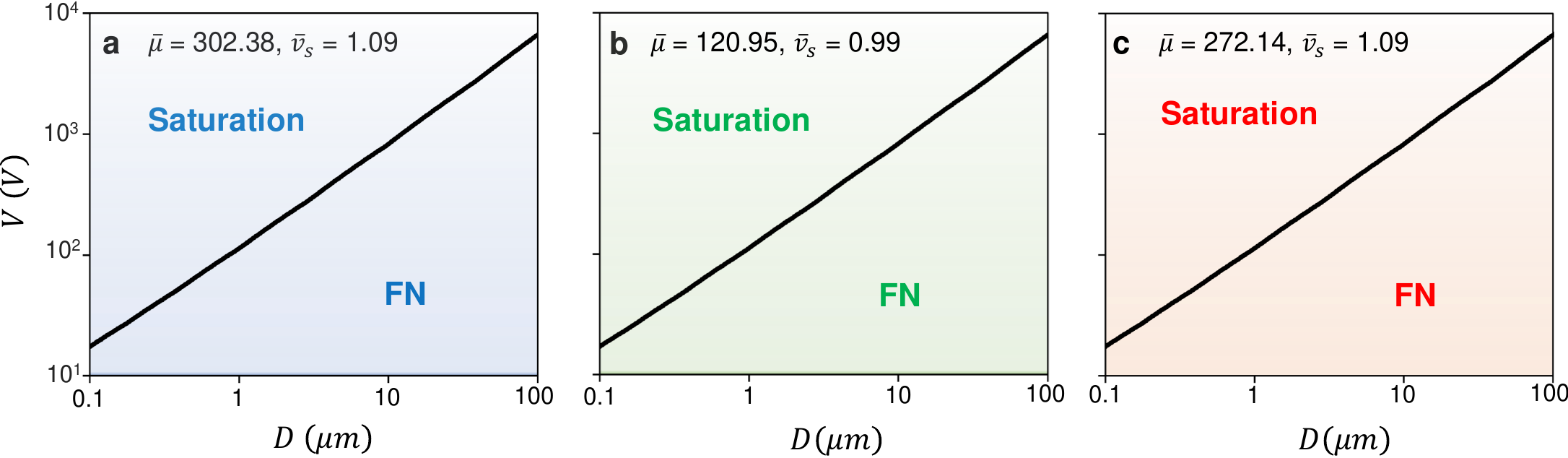}
        \caption{\(V - D\) nexus plots for three common wide bandgap (WBG) semiconductors: (a) gallium nitride (GaN), (b) 6H-SiC and (c) 4H-SiC silicon carbide (SiC) polytypes. The value of the work function used to normalise both the mobility and saturation velocity is 0.2 eV.}
        \label{fig: Figure 5}
\end{figure*}

The effect of $\Bar{v}_s$ is investigated in Figs. \ref{fig: Figure 4}(c) and \ref{fig: Figure 4}(d) at $\Bar{D} = 10^5$ and $10^7$, respectively.
At low $\Bar{v}_s$, SCLC$_\text{VS}$ is the only high-voltage process due to the severe carrier velocity limitation of $\Bar{v}_s$, while SCLC$_\text{MG}$ and/or SCLC$_\text{CL}$ only emerge when the velocity saturation effect is relaxed (i.e. larger $\Bar{v}_s$). 
The effect of carrier mobility is investigated in Figs. \ref{fig: Figure 4}(e) and \ref{fig: Figure 4}(f) at $\Bar{v}_s = 0.1$ and $\Bar{v}_s = 1000$, respectively. Contrary to the high-$\Bar{v}_s$ case [Fig. \ref{fig: Figure 4}(f)], the high-voltage transport is dominated by SCLC$_\text{VS}$ at low $\Bar{v}_s$. The SCLC$_\text{MG}$ can, however, be resurrected in low-mobility devices, in which the carrier collisional and scattering effects can momentarily forbid the carrier velocity to saturate at the intermediate-voltage regime. 

Finally, the transitional behaviours of three representative WBS, namely GaN, 6H-SiC and 4H-SiC, are shown in Figs. \ref{fig: Figure 5}(a) to \ref{fig: Figure 5}(c). As the transition boundary equation of FN-SCLC$_\text{VS}$ is $\Bar{v}_s$-dependent but $\Bar{\mu}$-independent  (see Fig. \ref{fig: Figure 3}), the $V$-$D$ diagram exhibits similar behaviours for all three WBS (whose $v_s$ values are similar). Here, FN transits directly into the velocity saturation regime without exhibiting any MG-type scaling, thus suggesting the dominance of SCLC$_\text{VS}$ over SCLC$_\text{MG}$, thus illustrating the importance of SCLC$_\text{VS}$, especially in the range of $10^2 \sim 10^4$ V which is relevant to the operating voltage of typical WBS-based power electronics with sub-$\mu$m channel length \cite{flack2016gan}.

In summary, we constructed a transitional transport model between field emission and space-charge-limited current with inclusion of the velocity saturation effect -- an important behaviour especially in devices composed of (ultra)wide bandgap semiconductors and insulators. Using nexus plots spanning various device/material parameter spaces, the rich transitional behaviours between various conduction mechanisms were revealed. Using GaN, 6H-SiC and 4H-SiC as illustrative examples of WBG semiconductors, we showed that the velocity-saturated space-charge-limited current is the dominant conduction mechanism at typical sub-$\mu$m device operating at $10^2 \sim 10^4$ V, thus unraveling the importance of velocity saturation effect. Finally, we remark that our model can be further generalized to include the effects of thermionic \cite{Murphy-Good}, quantum tunneling \cite{zhang2015scaling, zhou2023theoretical} and photo-induced emissions \cite{10.1063/1.2752122, 10.1116/1.1573664}, as well as unconventional electron emitters \cite{Chan-Chua, Chan-Ang, Ang-Cao, Ang-Chen}. 


\begin{acknowledgments}
\vspace{-1em} 
This work is supported by the Singapore Ministry of Education Academic Research Fund Tier 2 (Award No. MOE-T2EP50221-0019) and the SUTD Kickstarter Initiatives (SKI) (Award No. SKI 2021\_01\_12).
\end{acknowledgments}

\section*{AUTHOR DECLARATIONS}
\vspace{-1em} 
\subsection*{Conflict of Interest}
\vspace{-1em} 
The authors declare no conflict of interest.

\subsection*{Author Contributions}
\vspace{-1em} 
\textbf{Kok Wai Lee}: Investigation (equal); Formal analysis (equal); Visualization (equal); Writing – original draft (lead); Visualization (lead); Writing – review and editing (supporting). \textbf{Yee Sin Ang}: Conceptualization (lead); Investigation (equal); Funding acquisition (lead); Formal analysis (equal); Supervision (lead); Visualization (supporting); Writing – original draft (supporting); Writing – review and editing (lead).

\section*{DATA AVAILABILITY}
\vspace{-1em} 
The data that support the findings of this study are available
from the corresponding author upon reasonable request.


\providecommand{\noopsort}[1]{}\providecommand{\singleletter}[1]{#1}%

\end{document}